\documentclass[10pt]{iopart}
\usepackage{iopams,paralist,graphicx,hyperref,soul}

\newcommand{\ie}{i.e.\ }
\newcommand{\viz}{{\it viz.}\ }
\newcommand{\eg}{e.g.\ }

\newcommand{\ust}{u_\ast}
\newcommand{\ustt}{u_{\ast \mathrm{t}}}

\newcommand{\sal}{\mathrm{sal}}
\newcommand{\rep}{\mathrm{rep}}
\newcommand{\air}{\mathrm{air}}
\newcommand{\taug}{\tau_\mathrm{g}}
\newcommand{\taua}{\tau_\mathrm{a}}
\newcommand{\taut}{\tau_\mathrm{t}}
\newcommand{\rmm}{\mathrm{m}}
\newcommand{\rmc}{\mathrm{c}}
\newcommand{\Ei}{\mathrm{E}_1}
 
\begin{document}

\title{A two-species continuum model for aeolian sand transport}
\date{\today}

\author{M L\"ammel, D Rings and K Kroy}

\address{Institute for Theoretical Physics -- University of Leipzig\\
Postfach 100\,920, 04009 Leipzig, Gemrnay}
\ead{klaus.kroy@uni-leipzig.de}

\begin{abstract}
  Starting from the physics on the grain scale, we develop a simple
  continuum description of aeolian sand transport.  Beyond popular
  mean-field models, but without sacrificing their computational
  efficiency, it accounts for both dominant grain populations, hopping
  (or ``saltating'') and creeping (or ``reptating'') grains.  The
  predicted stationary sand transport rate is in excellent agreement
  with wind tunnel experiments simulating wind conditions ranging from
  the onset of saltation to storms.  Our closed set of equations thus
  provides an analytically tractable, numerically precise and
  computationally efficient starting point for applications addressing
  a wealth of phenomena from dune formation to dust emission.
\end{abstract}

\pacs{45.70.Mg, 47.27.T-, 47.55.Kf, 92.40.Gc}
% 92.40.Gc Hydrology and glaciology; cryosphere: Erosion and sedimentation; sediment transport
% 45.70.-n Granular systems (see also 05.65.+b Self-organized systems)
% 45.70.Mg Granular flow: mixing, segregation and stratification
% 47.27.T- Turbulent transport processes 
% 47.27.-i Turbulent flows
%47.55.Kf Particle-laden flows 
\submitto{\NJP}
\maketitle

\section{Introduction}
\label{sec:intro}
Wind-driven sand transport is the most noticeable process shaping the
morphology of arid regions on the Earth, Mars and elsewhere.  It is
responsible for the spontaneous creation of a whole hierarchy of
self-organized dynamic structures from ripples over isolated dunes to
devastating fields of shifting sands.  It also contributes
considerably to dust proliferation, which is a major determinant of
our global climate.  There is thus an urgent need for mathematical
models that can efficiently and accurately predict aeolian sand
fluxes. But the task is very much complicated by the complex turbulent
flow of the driving medium (e.g.\ streaming air) and the erratic
nature of the grain hopping it excites \cite{Bagnold1954}.  Yet, it is
by now well understood that this hopping of grains accelerated by the
wind has some characteristic structure \cite{Anderson1991a}.  Highly
energetic ``saltating'' grains make a dominant contribution to the
overall mass transport. When impacting on the sand bed, they dissipate
some of their energy in a complex process called splash, ejecting a
cloud of ``reptating'' grains \cite{Ungar1987,Anderson1987}.  A
snapshot of this splash would show a whole ensemble of trajectories
corresponding to a distribution of jump lengths from short, over
intermediate, to large hops.  However, grain-scale studies and
theoretical considerations have indicated that a reduced description
in terms of only two idealized populations (sometimes called saltons
and reptons) should indeed be able to provide a faithful
parametrization of the complex aeolian sand transport process and the
ensuing structure formation \cite{Anderson1991a,Andreotti2004}.  A
rule of thumb to say which grains in a real splash pertain to
idealized population of saltating or reptating grains is, whether or
not their jump height exceeds a threshold of about a few grain
diameters.  

A drawback of the two-species description has been that it is still
conceptually and computationally quite demanding.  For reasons of
simplicity and computational efficiency, many theoretical studies have
therefore chosen to reduce the mathematical description even further,
to mean-field type ``single-trajectory'' models \cite{Anderson1991}.
This may be justified for certain purposes, \eg for the mathematical
modelling of aeolian sand dunes, which are orders of magnitude larger
than the characteristic length scales involved in the saltation
process, and thus not expected to be very sensitive to the details on
the grain scale. Their formation and migration is thought to
predominantly depend on large-scale features of the wind field and of
the saltation flux, chiefly the symmetry breaking of the turbulent
flow over the dune and the delayed reaction of the sand transport to
the wind \cite{Kroy2002}.  Moreover, the reptating grains, although
they are many, are generally thought to contribute less to the overall
sand transport, because they have short trajectories and quickly get
trapped in the bed again.  Therefore, it seems admissible to
concentrate on the saltating particles.  On this basis, numerically
efficient models for one effective grain species that can largely be
identified with the saltating grain fraction, have been constructed.
The Sauermann model \cite{Sauermann2001} is a popular and widely used
example of such mean-field continuum models.  One-species models have
become a very successful means of gaining analytical insight into
\cite{Andreotti2002,Andreotti2002a,Sorensen2004,Duran2006,Fischer2008,Jenkins2010}
and to conduct efficient large-scale numerical simulations of
\cite{Schwammle2005,Duran2006a,Parteli2007,Parteli2007a,Duran2010} the
complex structure formation processes caused by aeolian transport.
The reduction to a single representative trajectory makes the
one-species models analytically tractable and computationally
efficient.  However, it is also responsible for some weaknesses both
concerning the way how the two species are subsumed into one
\cite{Andreotti2004} and how they feed back onto the wind
\cite{Duran2006}. These entail imperfections in the model predictions,
most noticeable a systematic overestimation of the stationary flux at
high wind speed (see \fref{fig:fluxlaws}, below).  Therefore, the
one-species models have been criticized for their lack of numerical
accuracy and internal consistency \cite{Andreotti2004,Andreotti2007}.
There are also a number of interesting phenomena that cannot be
quantitatively modelled within a one-species model, because they
specifically depend on one of the two species, or on their
interaction, or because the two species are not merely conceptually
but also physically different.

As three important representatives of such phenomena, we name dust
emission from the desert, ripples, and megaripples.  Dust is created,
exposed to the wind, and emitted by aeolian sand transport, and
saltating and reptating particles play quite different roles in these
processes \cite{Bagnold1954,Cahill1996}.  For example, fragmentation
processes might be driven by the bombardment of high-energy saltating
grains \cite{Kok2010a}, but not by the slower reptating grains.  In
contrast, the emission of dust hidden from the wind underneath the top
layer of grains in a sand bed could arguably be linked to the absolute
number of reptating grains dislodged from the bed.  The reptating
grain fraction and its sensitivity to the local slope of the sand bed
are, moreover, held responsible for the spontaneous evolution of
aeolian ripples
\cite{Bagnold1954,Anderson1987,Anderson1991a,Anderson1990,Prigozhin1999,%
  Hoyle1999,Valance1999,Yizhaq2004a,Andreotti2006,Manukyan2009}. But
the amount of reptating particles depends on the strength of the
splash caused by the saltating grains.  This interdependence of the
reptating and saltating species becomes even more transparent when the
two species correspond to visibly different types of grains.  This is
the case for megaripples, which have wavelengths in the metre range
and form on strongly polydisperse sand beds in a process accompanied
by a pronounced grain sorting \cite{Yizhaq2008,Milana2009,Yizhaq2012}.
The mechanism underlying their formation and evolution is still not
entirely understood, but it seems that the highly energetic
bombardment of small saltating grains drives the creep of the larger
grains armouring the ripple crests.  The mentioned grain-sorting
emerges due to the different hop lengths of the grains of different
sizes.

The demand for more faithful descriptions of aeolian sand transport
has recently spurred notable efforts to eliminate the deficiencies of
the one-species models \cite{Sorensen2004,Duran2006,Paehtz2012} or to
develop a numerically efficient two-species model
\cite{Andreotti2004}.  It also motivated the development of the model
described in detail below, as we felt that the problem has not yet
been cured at its root.  In our opinion, many of the amendments
proposed so far have targeted the symptoms rather than the disease by
invoking some {\it ad hoc} assumptions.  It was our aim to modify the
sand transport equations from bottom-up, based on a careful analysis
of the physics on the grain scale and including the feedback of the
two dissimilar grain populations on the turbulent flow.  In this way,
we derived a conceptually simple, analytically tractable, and
numerically efficient two-species model with only two phenomenological
fit parameters.  They serve to parametrize the complicated splash
process and take very reasonable values if the predicted flux law is
fitted to measured data.  Compared with the time when the first
continuum sand transport models were formulated, we could build on a
more detailed experimental and theoretical knowledge of the
grain-scale physics
\cite{Andreotti2004,Beladjine2007,Ammi2009,Valance2009,Kok2009,%
  Kang2011,Duran2011} and rely on more comprehensive empirical
information about the wind shear stress and sand transport to test our
predictions
\cite{Iversen1999,Rasmussen2005,Rasmussen2008,Creyssels2009}.
 
The plan of the paper is as follows. We first summarize some of the
pertinent basic notions of turbulent flows and aeolian sand transport
and introduce the two-species formalism. Then we implement the
two-species physics on the level of the sand flux in
\sref{sec:massbalance}, and on the level of the turbulent closure for
the wind in \sref{sec:wind}. In sections \ref{sec:kinetics} and
\ref{sec:transport-law}, we address the most interesting observables,
namely the speed and the average number of hopping grains, and the
saturated sand flux.  We finally compare our results with other models
and experiments in \sref{sec:discussion}, before concluding in
\sref{sec:conclusions}.

\section{Basic notions of aeolian sand transport and the two-species
  parametrization}
\label{sec:basics}
\begin{figure}[t]
  \centering
  \includegraphics[width=\linewidth]{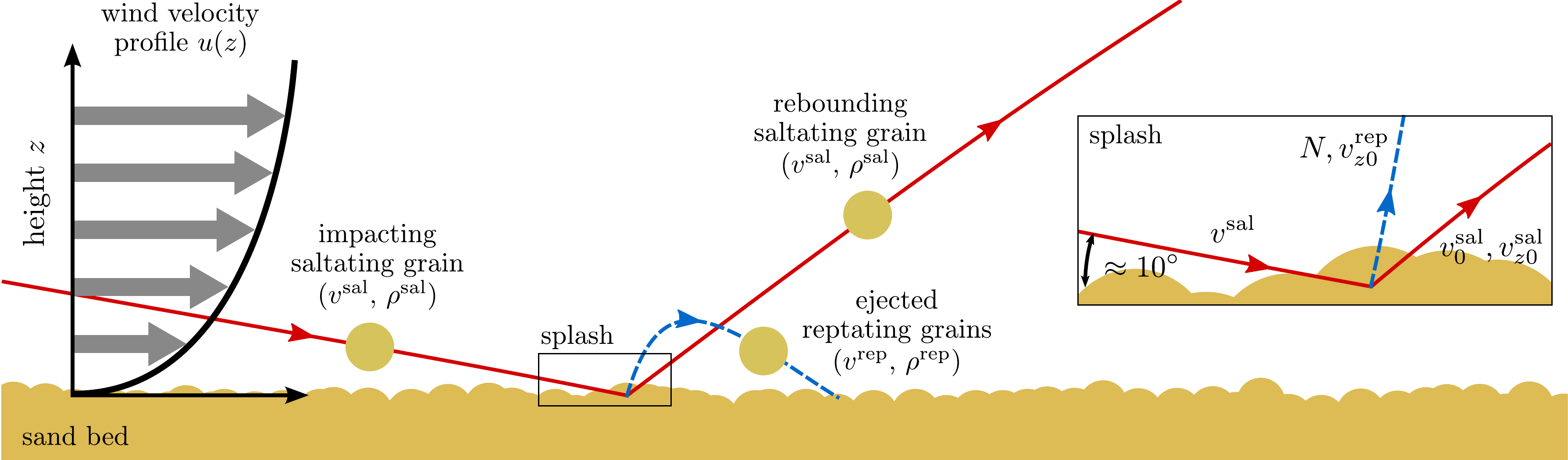}
  \caption{\textbf{The two-species picture of aeolian sand transport
      maps the ensemble of mobile grains onto two representative
      species.}  High-energy (``saltating'') grains accelerated by the
    wind (arrows in the wind-speed profile indicating the wind
    velocity) rebound upon impact and eject some low-energy
    (``reptating'') grains in a splash.  Sand transport is quantified
    in terms of the transport velocities, $v^\sal$ and $v^\rep$, and
    mean densities, $\rho^\sal$ and $\rho^\rep$, of the two species.
    The momentum balance of the splash process (inset, see
    \sref{sec:massbalance}) is effectively encoded in the impact
    velocity $v^\sal$, the horizontal and vertical rebound velocities
    of the saltating grain, $v^\sal_0$ and $v^\sal_{z0}$,
    respectively, and the number $N$ and velocity
    $v^\rep_{z0}=v^\rep_0$ of the ejected reptating grains (neglecting
    a small horizontal velocity component, for simplicity).}
  \label{fig:intro:sketch} 
\end{figure}

Our analysis of the two-phase flow of air and sand is similar, in
spirit, to the Sauermann model \cite{Sauermann2001}.  The Sauermann
model is a continuum or hydrodynamic model. Rather than with the
positions and velocities of individual grains it deals with
spatio-temporal fields of these quantities, \viz the local mass
density $\varrho(\vec x,t)$ and sand transport velocity $\vec v(\vec
x,t)$. It is of mean-field type since it reduces both fields to the
mean quantities $\rho(x,t)$, $v(x,t)$ for a single representative
trajectory, characterizing the conditions along the wind direction
(the $x$-direction), with the distribution of airborne grains in the
vertical direction (the $z$-direction) integrated out.  Note that
$\rho$ therefore is an area density, obtained from integrating the
volume density $\varrho$ over $z$. (The vertical grain distribution is
a crucial element of the modified turbulent closure, though, which
accounts for the feedback of the sand onto the wind.)  In this
contribution, we deal exclusively with the saturated sand flux, \ie
the sand flux along the wind direction over a flat sand bed and under
stationary wind conditions.  What we call the flux $q=\rho v$ is
actually a vertically integrated flux, or a sand transport rate
\cite{Sauermann2001}.  The central aim is to derive a constitutive
relation $q(\tau)$, giving the stationary flux at a given shear stress
$\tau$.

In the following, the strong mean-field approximation of the
one-species models is somewhat relaxed and reptating and saltating
particles with densities $\rho^\rep$, $\rho^\sal$ and transport
velocities $v^\rep$, $ v^\sal$ are treated separately.  The flux is
split up accordingly
\begin{equation}
  \label{eq:basics:flux-div}
   q = q^\rep + q^\sal  = \rho^\sal v^\sal + \rho^\rep v^\rep\,.
\end{equation}
Introducing the dimensionless mass fraction $\varphi \in [0,1]$ of
saltating grains relative to the total amount of mobile grains, the
(integrated) densities of the mobile grains can be written in the form
\begin{equation}
  \label{eq:basics:density-spec}
  \rho^\sal = \varphi \rho\quad \mathrm{and} \quad
  \rho^\rep = (1-\varphi) \rho \,.
\end{equation}
Accordingly, the flux balance and the relative contribution of
reptating and saltating grains to the flux read as
\begin{equation}
  \label{eq:basics:flux-div-weighted}
  \eqalign{
    q = \rho v= \rho \left[ \varphi v^\sal + (1-\varphi) v^\rep \right] \quad \mathrm{and} \\
    q^\rep/q^\sal = (\varphi^{-1}-1) v^\rep/v^\sal  \,.
  }
\end{equation}

The fluxes depend on the wind velocity field, which, in turn, depends
on the grain-wind feedback.  To make progress in this respect, the
(steady) shear stress $\tau$ in the saltation layer is split up into
two components carried by the airborne grains (``g'') and the air
(``a'') itself, respectively,
\begin{equation}
  \label{eq:basics:stressdecomp}
  \tau = \taug(z)+\taua(z) = \mathrm{const.}
\end{equation}
This idea, introduced by Owen \cite{Owen1964}, which amounts to an
effective two-fluid representation of the wind and the airborne sand,
was used in many analytical models
\cite{Andreotti2004,Sauermann2001,Sorensen2004,Raupach1991} and
confirmed by numerical simulations
\cite{Anderson1991,Kok2009,Kang2011,Duran2011,McEwan1991}.  In the
two-species model, we moreover keep track of the two species of mobile
grains in \eref{eq:basics:stressdecomp}.  Similar to the flux division
introduced in \eref{eq:basics:flux-div}, we further split up
the grain-borne shear stress
\begin{equation}
  \label{eq:basics:grainstress-div}
  \taug(z) = \taug^\rep(z) + \taug^\sal(z) \,.
\end{equation}
Concerning the feedback of the grains on the wind, it turns out that
the wind velocity profile $u(z)$ is predominantly determined by the
reptating grain fraction and thus by the functional form of
$\taug^\rep(z)$, while the number and trajectories of the saltating
grains hardly affect the wind profile at all.  Physically, this makes
sense, since the saltating particles form a very disperse gas compared
with the dense reptating layer.  Also note that the reptating grains,
which lose all their momentum upon impact, are responsible for the
momentum loss of the flow field.  The wind profile may thus be
computed in a reduced one-species scheme, as explained in
\sref{sec:wind} and \ref{sec:app:2spec-wind}.

Under steady conditions, a common simplifying assumption is that the
airborne stress near the ground ($z=0$) can be identified with the
threshold shear stress $\taut$ required to mobilize grains from the
ground \cite{Owen1964},
\begin{equation}
  \label{eq:basics:owen-assumption}
  \taua(0) \approx \taut \,.
\end{equation}
If it were larger or smaller, an increasing or decreasing number of
grains would be mobilized, respectively, resulting in an unsteady
flow.  This idea is indeed supported by empirical observations, which
find that the number of ejected grains increases (almost linearly)
with the impact speed, not their ejection velocity or jump height
\cite{Beladjine2007,Ammi2009,Rioual2000,Rioual2003}. The feedback of
the grains on the wind thus essentially fixes the air shear stress at
the ground (and, if $\tau \approx \taut$, also above) to the threshold
value $\taut$. To be precise, $\taut$ is called the impact threshold,
to emphasize that it is easier to lift grains from the splash cloud
than to lift completely immobile grains from the ground
\cite{Bagnold1954}.  But, for the sake of simplicity, we do not bother
to make this distinction here, nor do we speculate about possible
small deviations of $\taua(0)$ from $\taut$ under steady conditions
\cite{Duran2006,Paehtz2012}.  Combining \eref{eq:basics:stressdecomp}
and \eref{eq:basics:owen-assumption}, we may express the shear stress
contributed by the grains near the ground as
\begin{equation}
  \label{eq:basics:taug-ground}
  \taug(0) = \taug^\rep(0) + \taug^\sal(0) =\tau-\taut \,.
\end{equation}
Both stress contributions are defined as the product of the vertical
sand flux $\phi^i$ and the horizontal velocity difference $v^i-v_0^i$
upon impact, $\taug^i(0)= (v^i-v_0^i)\phi^i$.  The vertical sand flux
$\phi^i$ is given by the horizontal (height integrated) flux $q^i$
divided by the grains' mean hop length, $\phi^i \propto q^i = \rho^i
v^i$.  Here, $i$ represents the species indices ``$\rep$'' or
``$\sal$''.  Following Sauermann \etal \cite{Sauermann2001}, we assume
parabolic grain trajectories with initial horizontal and vertical
velocity components $v_0^i$ and $v_{z0}^i$, respectively, and obtain
\begin{equation}
  \label{eq:basics:taug-ground-parabola}
  \phi^i = g \rho^i /(2 v_{z0}^i) \,, \quad \taug^i(0) = g \rho^i (v^i-v_0^i)/(2 v_{z0}^i) \,,  \quad i=\rep,\,\sal\,.
\end{equation} 
Since the average ejection angle of the reptating particles is known
to be about $80^\circ$ \cite{Beladjine2007}, $v_0^\rep\approx 0$ and
we identify the total ejection speed of the reptating grains with its
vertical component $v^\rep_{z0}$.  The impact angle of saltating
grains is known to be almost independent of the wind strength and to
take typical mean values of about $10^\circ$ \cite{Rice1995}, so that we
can identify the total impact speed with the horizontal speed $v^\sal$
of the impacting particle, see \fref{fig:intro:sketch}.

In order to derive the sought-after constitutive equation, we have to
understand the balance of the two species of mobile grains.  If one
considers the vertical fluxes $\phi^\sal$ and $\phi^\rep$ rather than
the horizontal fluxes $q^\sal$ and $q^\rep$, one has
\begin{equation}
  \label{eq:basics:vertical-fluxes}
  \phi^\rep = N \phi^\sal \,,
\end{equation}
with the average number $N$ of reptating grains ejected by an
impacting saltating grain.  Together with the first relation in
\eref{eq:basics:taug-ground-parabola}, we then have
\begin{equation}
  \label{eq:basics:weightfct}
  \varphi^{-1}  = 1 + N v^\rep_{z0}/v^\sal_{z0} \,.
\end{equation}
Using also the second relation in
\eref{eq:basics:taug-ground-parabola} and remembering that we
dismissed the horizontal component of the ejection velocity, we can
thus write the overall density of mobile grains in the form
\begin{equation}
  \label{eq:basics:overall-density}
  \rho = \frac{2 \varrho_\air}{g} \frac{\ust^2-\ustt^2}{\varphi
    (v^\sal-v^\sal_0)/v^\sal_{z0} + (1-\varphi) v^\rep/v^\rep_{z0}} \,.
\end{equation} 
As often found in the literature, we have employed the notion of the
shear velocity or friction velocity $\ust$, defined by $\tau \equiv
\varrho_\air \ust^2$, as a more intuitive measure of the shear stress
$\tau$ here. The velocity ratios in the denominator of
\eref{eq:basics:overall-density} are supposedly only very weakly
dependent on the wind strength. The first one is recognized as an
effective restitution coefficient
\begin{equation}
  \label{eq:basics:alpha}
  \alpha \equiv v^\sal_{z0}/(v^\sal-v^\sal_0)
\end{equation} 
of the saltating grains \cite{Sauermann2001}.  Because of the
complexity of the splash process, which makes first-principle
quantitative estimates of $\alpha$ forbiddingly complex, we suggest to
treat it as a phenomenological fit parameter.  We do however provide a
reasonable theoretical estimate of the ratio $v^\rep/v^\rep_{z0}$,
based on the trajectory of a vertically ejected grain driven by the
wind, in \sref{sec:kinetics} and \ref{sec:app:reptation-velocity}.
Anticipating the result $v^\rep/v^\rep_{z0} = 0.7 (\sigma \nu_\air
g)^{1/3} /\ustt$, we can rewrite the density of the transported sand
in the more explicit form
\begin{equation}
  \label{eq:basics:overall-density-alpha}
  \rho = \frac{2 \varrho_\air}{g} \frac{\ust^2-\ustt^2}{\varphi/\alpha
    + 0.7 (1-\varphi) (\sigma \nu_\air  g)^{1/3}/\ustt} \,.
\end{equation} 
Here $\nu_\air$ denotes the kinematic viscosity of air, and since the
grain-air density ratio $\sigma \gg 1$ under terrestrial conditions,
we do not bother to distinguish between $\sigma$ and $\sigma-1$ here
and in the following.  Note that on top of the explicit wind strength
dependence of $\rho$ via the numerator, there is an implicit, yet
undetermined one via $\varphi$.

In the following, we develop the sought-after ``second generation''
transport law $q(\ust)$, based on the grain-scale physics. Some
empirical input is employed to fix certain phenomenological
coefficients summarized in \tref{tab:parameters}.  Our result (see
\tref{tab:transport-law:one-spiecies-limit:fluxlaws}) accounts for
essential elements of the aeolian sand transport process beyond the
single-trajectory approximation, and turns out to be in remarkable
agreement with wind tunnel measurements.

\section{Implementing the two-species framework}
\label{sec:2spec}
The two-species framework is implemented in three steps.  First, we
deal with the mass balance between the species, then with their
feedback onto the wind and their resulting transport velocities,
before we finally arrive at an improved stationary transport law
$q(\ust)$.

\subsection{Sand density: two-species mass balance}
\label{sec:massbalance}
We first estimate the partition of the grain population into a
saltating and a reptating fraction from several evident assumptions
for the splash process, based on fundamental physical principles.  In
our formalism, a saltating grain always rebounds upon impact and
ejects $N$ reptating grains, while a certain fraction of its kinetic
energy and momentum is dissipated in the sand bed.  We further assume
that every reptating grain performs only a single hop after which it
remains trapped in the sand bed.  These mild approximations for the
stationary sand transport reduce the number of free parameters of our
model compared with more elaborate descriptions \cite{Andreotti2004}.

As argued, e.g., by Kok and Renno \cite{Kok2009}, the relevant constraint
limiting the number $N$ of ejected grains per impact is given by the
conservation of the grain-borne momentum rather than the energy.  In
writing the momentum balance, we make a physically plausible
linear-response approximation, namely that all terms scale linearly in
the velocity $v^\sal$ of the impacting grain.  This amounts to
assuming a wind-strength-independent scattering geometry, so that we
can add $z$- and $x$-components up to constant geometric factors that
are later absorbed in phenomenological coefficients.  Including also
the bed losses proportional to $v_\mathrm{bed}$ in this vein, we have
\begin{equation}
  \label{eq:massbalance:momentum-balance}
  v^\sal = v^\sal_0  + v^\sal_{z0} + N v^\rep_{z0} +v_\mathrm{bed} \,.
\end{equation}
For the first two terms on the right-hand side of
\eref{eq:massbalance:momentum-balance}, the proportionality to $v^\sal$ is already
implicit in the definition \eref{eq:basics:alpha} of the restitution
coefficient $\alpha$. There is an important subtlety concerning the
last two terms, however.  Namely, the ejected grains have to gain
enough momentum to jump over neighbouring grains in the bed in order
to be counted as mobile grains.  But they also should not themselves
eject other grains upon impact; otherwise we would have to count them
among the saltating fraction.  Hence, the impact velocity $v^\rep$ and
the ejection velocity $v^\rep_{z0}$ of the reptating grains are
tightly constrained and cannot, by definition, be strongly dependent
of the impact velocity of the saltating grain or the wind strength.
Within experimental errors, observations of the collision process of
saltating grains by Rice \etal \cite{Rice1995} are indeed in
reasonable agreement with a constant $v^\rep_{z0}$ of the order of
$\ustt$.  In fact, they found the horizontal component of the ejection
velocity to be independent of $v^\sal$, and only a slight increase of
the vertical component with $v^\sal$.  A compelling confirmation of
this observation was recently obtained by better controlled laboratory
experiments, in which PVC beads were injected at an impact angle of
$10^\circ$ onto a quiescent bed of such beads
\cite{Beladjine2007,Ammi2009}.  There are two important consequences
of the constraints on $v^\rep_{z0}$.  Firstly, $N$ should grow
linearly with $v^\sal$.  Secondly, since the momentum of the
impacting grain is only partially transferred to the ejected grains, a
critical impact velocity $v^\sal = v^\sal_\rmc\gg v^\rep_{z0}$, has to
be overcome to mobilize any grains at all.  It can be interpreted as a
constant offset in $v_\mathrm{bed}$, which does not scale with
$v^\sal$.  According to the collision experiments
\cite{Beladjine2007,Ammi2009}, one can take $v^\sal_\rmc\approx
40\sqrt{g d}\approx 9 \ustt$, where we used Bagnold's estimate
$\ustt^2\approx 0.01 \sigma g d$ \cite{Bagnold1954}.  A concise
summary of the empirical input entering our model is provided in
\tref{tab:parameters}.

Summarizing the above discussion, we can rewrite the momentum balance
\eref{eq:massbalance:momentum-balance} as
\begin{equation}
  \label{eq:massbalance:ejection-number}
   N v^\rep_{z0} \propto v^\sal - v^\sal_\rmc \,,
\end{equation}
where $v^\sal >v^\sal_\rmc$ is tacitly assumed to hold throughout our
discussion, and the omitted factor of proportionality should be
insensitive to the wind strength.  Observations of saltating grains
\cite{Rice1995}, model collision experiments
\cite{Beladjine2007,Ammi2009,Rioual2000,Rioual2003}, particle dynamics
simulations \cite{Anderson1991,Anderson1988} and reduced numerical
models, such as the binary collision scheme proposed by Valance and
Crassous \cite{Valance2009}, all support this affine relation of the
number of ejected grains, which is the cornerstone of our two-species
mass balance.  To fix the omitted factor in
\eref{eq:massbalance:ejection-number} and make contact with
\eref{eq:basics:weightfct}, we write
\begin{equation}
  \label{eq:massbalance:ejection-number-2}
  N v^\rep_{z0} = (\eta/ \alpha) \left( 1- v^\sal_\rmc/v^\sal \right )v^\sal_{z0} \,.
\end{equation}
The constant $\eta$, which serves as the second free fit parameter of
our model, determines together with $v^\sal_\rmc$ how the momentum
lost by a rebounding grain is distributed between the bed and the
ejected particles.  (Details of the phenomenological values of
$\alpha$ and $\eta$ obtained by fitting the complete model to
the experimental flux data are given below.)  Comparison with
\eref{eq:basics:weightfct} now yields an explicit expression of the
mass fraction
\begin{equation}
  \label{eq:massbalance:weightfct-alpha}
  \varphi^{-1} = 1 + (1 - v^\sal_\rmc/v^\sal)\eta/\alpha 
\end{equation}
in terms of $v^\sal$, alone. Thereby, the problem of specifying the
stationary mass balance between saltating and reptating grains has
been completely reduced to the task of finding the stationary
transport velocity $v^\sal$ of the saltating particles as a function
of the wind strength.

\subsection{Wind velocity field: two-species turbulent closure}
\label{sec:wind}
\begin{figure}[t]
  \centering
  \includegraphics[width=0.6\linewidth]{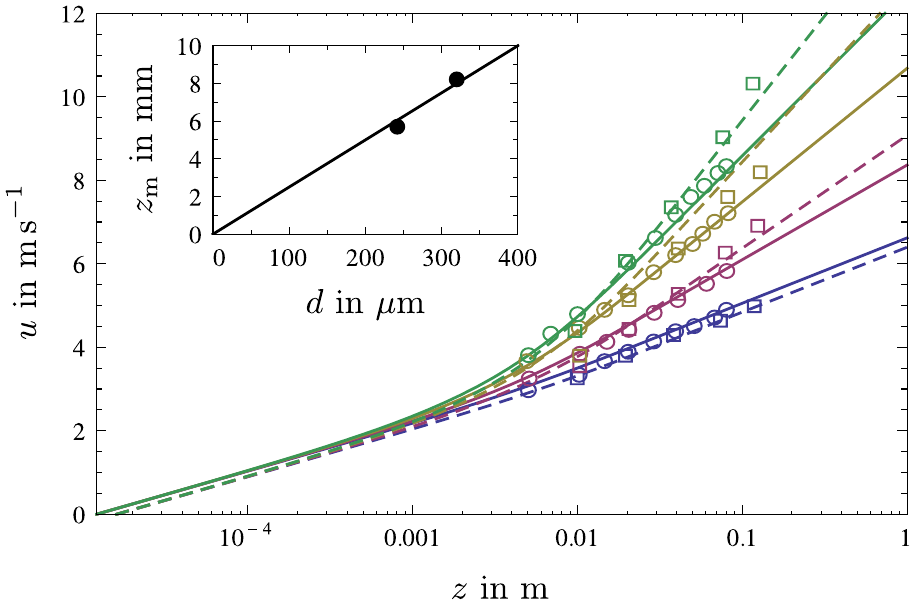}
  \caption{{\bf The wind speed $u$ at height $z$.} Theory, equation
    \eref{eq:wind:windprofile} (curves), compared with the measured data
    (symbols) for grains of diameter $d=242\,\mathrm{\mu m}$ (solid
    curves, circles) and $d=320\,\mathrm{\mu m}$ (dashed curves,
    boxes) and various shear velocities ($\ust = 0.27$, $0.39$, $
    0.56$, $0.69\, \mathrm{m\, s^{-1}} $ and $\ust = 0.27$, $0.47$, $
    0.74$, $0.87\;\mathrm{m\, s^{-1}}$ from bottom to top for
    $d=242\;\mathrm{\mu m}$ \cite{Rasmussen2008} and $d=320\;
    \mathrm{\mu m}$ \cite{Rasmussen2005}, respectively); as global fit
    parameters we used the roughness length ($z_0 = d/20$), the
    threshold shear velocity ($\ustt=0.19\;\mathrm{m\,s^{-1}}$,
    $0.20\;\mathrm{m\,s^{-1}}$ for the finer/coarser sand) and the
    mean saltation height $z_\rmm$, which is compared with
    \eref{eq:wind:zm} in the inset.}
  \label{fig:windprof}
\end{figure}

Before we can work out the transport velocities of the two species of
grains, we need to know the height-dependent wind speed $u(z)$ and how
it is affected by the presence of the airborne grains. Unfortunately,
this leads us back to the question of the grain densities that we just
delegated to the calculation of the grain velocities.  In the past, an
elaborate self-consistent calculation has been avoided by anticipating
the resulting form of the wind profile on the basis of empirical
observations.  In the Sauermann model, for instance, an exponential
vertical decay of the grain-borne shear stress is imposed,
\begin{equation}
  \label{eq:wind:taug-vs-z}
  \taug(z) = \taug(0) \exp(-z/z_\rmm) \,,
\end{equation}
in good agreement with grain-scale simulations
\cite{Anderson1991,Kok2009,Kang2011,McEwan1991}. Since the airborne
stress $\taua(z)$ follows from this via
\eref{eq:basics:stressdecomp}, it reduces the task to the problem
of finding the mean saltation height $z_\rmm$.  In
\ref{sec:app:2spec-wind}, we present a refined version of this
approach, adapted to the two-species approach.  It accounts for the
contributions of the different species to the grain-borne shear
stress, separately, and also for their considerably different
trajectories.  However, as we detail in the appendix, the whole
exercise yields essentially identical results as
\eref{eq:wind:taug-vs-z}, which can in fact be interpreted as the
pre-averaged two-species expression.  This lends further support to
the general form of \eref{eq:wind:taug-vs-z}, recommending it as
a suitable basis for our further discussion of the two-species model.

From \eref{eq:wind:taug-vs-z}, we get, via
\eref{eq:basics:stressdecomp}, the wind speed profile $u(z)$ by
integrating Prandtl's turbulence closure
\begin{equation}
  \label{eq:wind:turbulence-closure}
  \taua(z) = \varrho_\air \kappa^2 z^2 [\partial_z u(z)]^2 \,.
\end{equation}
To proceed analytically, we use a modified secant approximation
similar to what was proposed by S{\o}rensen \cite{Sorensen2004}.  As
shown in \ref{sec:app:closure}, the closure equation then
becomes
\begin{equation}
  \label{eq:wind:closure-approx}
  \partial_z u \approx \frac{\ust}{\kappa z} \left[ 1 - ( 1 -
    \ustt/\ust) \rme^{-z/z_\rmm} \right] \,.
\end{equation}  
Upon integration from the roughness length $z_0\ll z_\rmm$, defined by
$u(z_0)=0$, this yields the explicit result
\cite{Sorensen2004,Duran2006}
\begin{equation}
  \label{eq:wind:windprofile}
  u(z) =  \frac{\ust}{\kappa}\ln(z/z_0) - \frac{\ust-\ustt}{\kappa}[\Ei (z_0/z_\rmm) - \Ei(z/z_\rmm)] \,,
\end{equation}
with the integral exponential $\Ei(z)\equiv\int_z^\infty\!  \rmd
x\,\rme^{-x}/x$. For small wind speeds $\ust\to \ustt$
($\ust\geq\ustt$), the usual logarithmic velocity profile is
recovered. Inside the saltation layer, a universal asymptotic wind
velocity field
\begin{equation}
  \label{eq:wind:universal}
  u(z) \sim \frac{\ustt}{\kappa} \ln(z/z_0) \qquad (z_0 \lesssim z \ll z_\rmm)
\end{equation}
emerges, which is independent of the wind strength outside the
saltation layer. This rationalizes the convergence of the wind
profiles for different wind strengths, found in wind tunnel
measurements, without the {\it ad hoc} assumption \cite{Duran2006} of
a debatable focal point \cite{Li2004}, which resulted in the
prediction of an unphysical negative flux $q(\ust)$ at high wind
speeds \cite{Duran2006}.  A direct comparison of our above prediction
for $u(z)$ with wind tunnel data from
\cite{Rasmussen2005,Rasmussen2008} can be found in
\fref{fig:windprof}.  From these data (available for two grain sizes
$d$) we determine the mean saltation height $z_\rmm$.  Past attempts
to relate $z_\rmm$ to $d$ invoked a new undetermined length scale
depending solely on atmospheric properties \cite{Andreotti2008}.
However, the results of our fit rather support the simpler (and more
natural) linear relation
\begin{equation}
  \label{eq:wind:zm}
  z_\rmm \approx 25 d
\end{equation}
(right panel of \fref{fig:windprof}).  This is in line with recent
analytical and numerical work \cite{Paehtz2012}, albeit possibly with a
somewhat different numerical factor dependent on specific model
assumptions.

\subsection{Transport kinetics: two-species transport velocities}
\label{sec:kinetics}
With the wind speed profile $u(z)$ at hand, we can now determine the
characteristic stationary transport velocity $v^\sal$ of the saltating
sand fraction from a standard force-balance argument.  We evaluate the
wind speed at a characteristic height $z^\sal$, at which the fluid
drag on the grains is then balanced with the effective bed friction,
in the usual way (for a detailed derivation and discussion, see
\cite{Sauermann2001,Duran2006}).  Namely, the drag force on a volume
element of the saltation cloud (the force per mass acting on a single
saltating grain times the density $\rho^\sal$ of the saltation cloud)
is balanced with $\taug^\sal(0)$.  This yields the transport velocity
of the saltating sand fraction
\begin{equation}
  \label{eq:kinetics:saltation-velocity}
  v^\sal = u \! \left(z^\sal\right) - v^\sal_\infty \,,
\end{equation}
with the estimate \cite{Duran2006,Jimenez2003}
\begin{equation}
  \label{eq:kinetics:terminal-velocity}
  v^\sal_\infty \approx \frac{ \sqrt{\sigma d g/\alpha} }{ 1.3 + 41 \, \nu_\air / \sqrt{\sigma d^3 g / \alpha} }
\end{equation}
for the terminal steady-state relative velocity $u-v^\sal$.  Formally,
$v^\sal_\infty$ is equal to the settling velocity asymptotically
reached by a grain freely falling in air (of kinematic viscosity
$\nu_\air$ and with grain--air density ratio $\sigma$), with a
rescaled gravitational acceleration $g \mapsto g/(2\alpha)$.

\begin{figure}[t]
  \centering
  \includegraphics[width=0.6\linewidth]{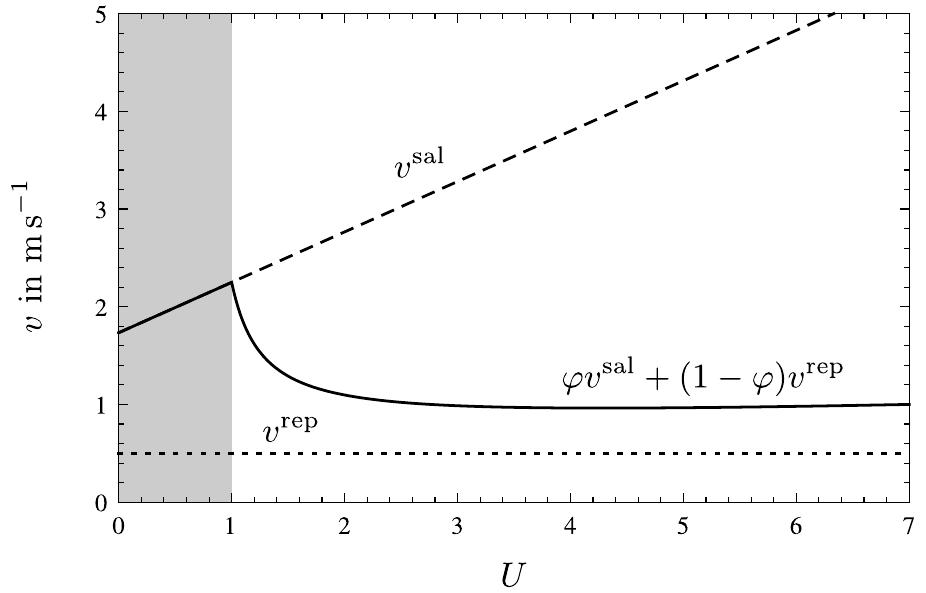}
  \caption{\textbf{The transport velocities of the saltating (dashed)
      and reptating (dotted) sand fraction versus the rescaled shear
      velocity $U\equiv \ust/\ustt$}, as predicted by
    \eref{eq:kinetics:saltation-velocity} and
    \eref{eq:kinetics:reptation-velocity}, respectively.  The solid
    line represents the average transport velocity in a reduced
    effective single-trajectory description.}
  \label{fig:kinetics:velocities}
\end{figure}

We fix the $\ust$-independent parameter $z^\sal$ by the plausible
assumption that the splash at the (impact) threshold wind speed
$\ustt$ dies out, corresponding to $v^\sal(\ust=\ustt) = v^\sal_\rmc$.
The underlying physical picture is that it is the splash that keeps
the saltation process going, even below the aerodynamic entrainment
threshold, because the reptating particles compensate for rebound
failures \cite{Andreotti2004}.  Accordingly, using
\eref{eq:kinetics:saltation-velocity} with the logarithmic wind field,
equation \eref{eq:wind:universal}, we find that
\begin{equation}
  \label{eq:kinetics:zsal}
  z^\sal=z_0 \, \exp \! \left(\kappa \frac{v^\sal_\rmc + v^\sal_\infty}{\ustt} \right) \,. 
\end{equation}
In \ref{sec:app:reptation-velocity}, we use a similar argument to derive
the mean reptation velocity, which turns out to be almost independent of
the grain size $d$.  It can thus be approximated by the constant
\begin{equation}
  \label{eq:kinetics:reptation-velocity}
  v^\rep \approx 0.7 (\sigma \nu_\air  g)^{1/3} \,,
\end{equation}
for most practical purposes.  This relation corresponds to a
$\ust$-independent reptation length $2 v^\rep v^\rep_{z0}/g = 0.14
[\sigma^5 \nu_\air^2 d^3 / g ]^{1/6}$ in the centimetre range, while
the saltation length $2 v^\sal v^\sal_{z0}/g \propto \ust^2/g$ is
quadratic in the shear velocity $\ust$ (see
\fref{fig:kinetics:velocities}) and of the order of decimetres
\cite{Bagnold1954,Nalpanis1993,Almeida2008}.

In \fref{fig:kinetics:velocities}, the transport velocities $v^\sal$
and $v^\rep$ for the two species and the mean velocity $v = \varphi
v^\sal + (1-\varphi) v^\rep$ are plotted over the shear velocity
$\ust$.  An important result is that the species-averaged transport
velocity $v$ is nearly constant over a broad range of wind strengths.
This is consistent with the fundamental assumption on which the model
is based, namely that the number of mobilized grains is sensitive to
the wind strength, but their overall transport kinetics is not.

\subsection{Stationary sand flux: two-species transport law}
\label{sec:transport-law}
The stationary flux law is now obtained by collecting the above
results for the saturated density $\rho$ from
\eref{eq:basics:overall-density}, the saltation fraction $\varphi$
from \eref{eq:basics:weightfct}, the transport velocities
\eref{eq:kinetics:saltation-velocity} with the wind velocity
\eref{eq:wind:windprofile}, and inserting them into
\eref{eq:basics:flux-div-weighted}. We introduce reduced variables,
measuring the shear velocity $\ust$ in terms of the impact threshold
$\ustt$ and the saturated flux $q$ in units of $\varrho_\air
\ust^3/g$,
\begin{equation}
  \label{eq:transport-law:scaling-fucntion}
  U\equiv \ust/\ustt \,, \qquad Q\equiv q g/(\varrho_\air \ust^3) \,.
\end{equation}
The result for our two-species stationary flux relation then takes the
form
\begin{equation}
  \label{eq:transport-law:flux}
  Q = \left(1-U^{-2}\right) \frac{\alpha_0U 
    + \beta_0  - \gamma_0 U^{-1}}{a U - b} \,,
\end{equation}
with the coefficients $\alpha_0$, $\beta_0$, $\gamma_0$, $a$ and $b$
depending on the two free parameters $\alpha$ and $\eta$, as
summarized in \ref{sec:app:coeff}.  It goes without saying that this
result pertains to $U>1$ and that $Q(U<1)\equiv 0$.  Except for $b$,
which becomes negative for small $\eta$, and in particular for
$\eta=0$, all coefficients are positive.  But since
$\gamma_0<\alpha_0+\beta_0$ and $b<a$, the flux always remains
positive.

The positivity of the flux in \eref{eq:transport-law:flux} guarantees
physically reasonable results, even under transient wind conditions,
as occurring, for example, in sand dune simulations.  The extension of
the present discussion to non-stationary conditions is a major task
for future investigations, in particular with regard to the saturation
length, \ie the characteristic length scale over which the system
relaxes towards the steady state \cite{Sauermann2001}.  Its derivation
within our two-species approach is of conceptual interest, since the
wind strength dependence of the saturation length has recently been
the subject of debates (see, e.g., \cite{Andreotti2007}).

\section{Discussion}
\label{sec:discussion}
Before we determine the free parameters $\alpha$ and $\eta$ by
comparing \eref{eq:transport-law:flux} with experimental data, we
first discuss the effect of our two-species parametrization on the two
related transport modes---saltation and reptation---and present two
different single trajectory reductions of the two-species model.
Subsequently, we compare these reduced schemes to single-species
transport laws that have previously been proposed in the literature.

\subsection{The two-species flux balance}
\label{sec:discussion:transport-law:flux-balance}
\begin{figure}[t]
  \includegraphics[width=0.48\linewidth]{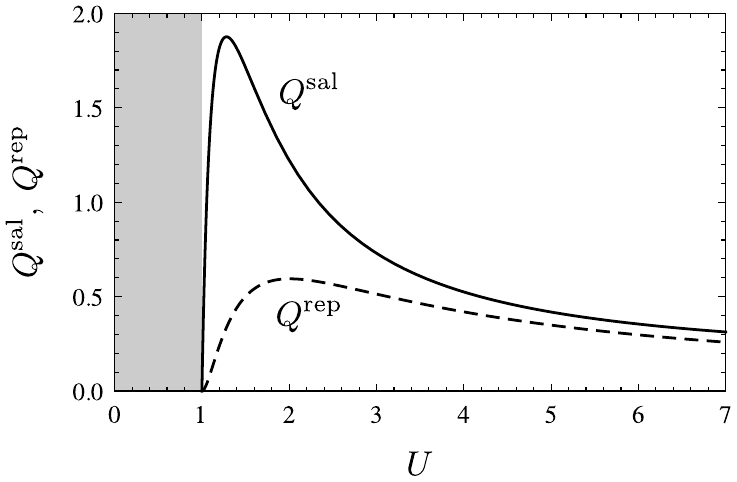}
  \hfill
  \includegraphics[width=0.48\linewidth]{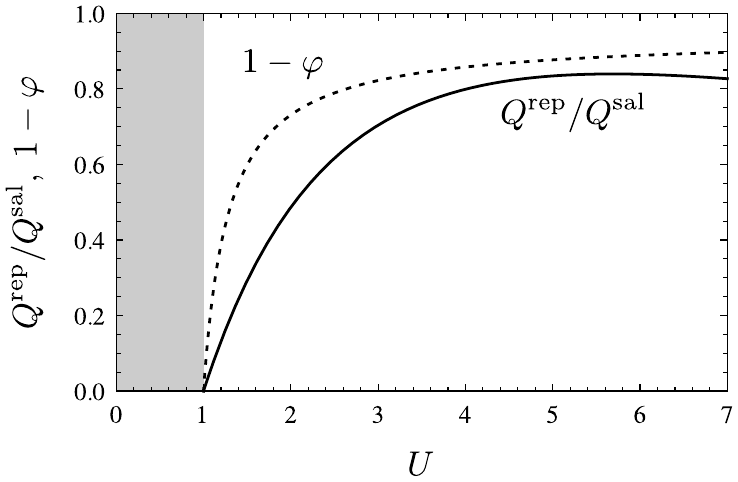}
  \caption{\textbf{The rescaled fluxes} $Q^\sal \equiv q^\sal
    g/(\varrho_\air \ust^3)$ and $Q^\rep \equiv q^\rep g/(\varrho_\air
    \ust^3)$ of the saltating and reptating grains, respectively,
    versus the rescaled shear velocity $U\equiv \ust/\ustt$
    (parameters: $\alpha=0.65$ , $\eta=9$, as determined in
    \sref{sec:experiments}; grain size $d=250\,\mathrm{\mu m}$).
    {\bf Left:} direct comparison of $Q^\sal$ (solid line) and
    $Q^\rep$ (dashed line).  {\bf Right:} the flux ratio $Q^\rep /
    Q^\sal = q^\rep / q^\sal$ (solid line) compared with the mass
    fraction $1 - \varphi$ of reptating grains (dashed line).}
  \label{fig:species-fluxes}
\end{figure}
Our derivation of the overall sand flux $Q$, given in
\eref{eq:transport-law:flux}, also yields the partial fluxes $Q^\sal
\equiv q^\sal g/(\varrho_\air \ust^3)$ and $Q^\rep \equiv q^\rep
g/(\varrho_\air \ust^3)$ of the saltating and reptating species,
respectively. Their dependence on the wind velocity is illustrated in
\fref{fig:species-fluxes}.  It is frequently argued in the literature
that the saltating grains dominate the sand transport, for they
perform large jumps, while the reptating sand fraction might be
negligible in terms of mass transport.  Our model basically confirms
this view for moderate winds, not too far above the impact threshold.
However, it predicts that both species contribute almost equally to
the total flux, under strong winds.  The shift in the mass balance
$\varphi$ towards reptation (\fref{fig:species-fluxes}, right plot)
reflects the dependence of the number of ejected grains on the wind
strength, which follows from the splash-impact relation
\eref{eq:massbalance:ejection-number} if one inserts the transport
velocities of the individual species, as given in
\fref{fig:kinetics:velocities}.  The variable contribution of the two
species to the overall sand flux is a result that cannot be obtained
from the conventional single-species models, but should be of interest
for some of the more advanced applications mentioned in the
introduction.

\subsection{One-species limits}
\label{sec:discussion:one-species-limits}
\begin{table}[b]
  \caption{Scaling functions $f(U)$ for stationary aeolian sand
    transport relations of the form $Q(U) =
    \left(1-U^{-2}\right)f(U)$.  Note that the coefficients of the
    various models cannot be identified even if represented by the
    same symbol. Explicit analytical expressions for the coefficients 
    $\alpha_0$, $\beta_0$, $\gamma_0$, $a$ and $b$ of the two-species 
    model can be found in \ref{sec:app:coeff}.
  }
  \label{tab:transport-law:one-spiecies-limit:fluxlaws}
  \begin{indented}
    \renewcommand{\arraystretch}{2} \setlength{\tabcolsep}{5pt}
  \item[] \begin{tabular}{@{}l l}
      \br
      This work & $\displaystyle \frac{\alpha_0U + \beta_0  - \gamma_0 U^{-1}}{a U - b}$ \\
      This work, one-species limit & $\displaystyle a - b U^{-1}$ \qquad $(b<0)$ \\
      Sauermann \etal 2001 \cite{Sauermann2001} & $a \sqrt{1+ \alpha_0 U^{-2}} + b U^{-1}$  \\
      S{\o}rensen 2004 \cite{Sorensen2004} & $a + b U^{-1} + c U^{-2}$ \\
      Dur{\'a}n and Herrmann 2006 \cite{Duran2006} & $a - b U^{-1}$ \qquad $(a,b<0)$ \\
      P\"ahtz \etal 2011 \cite{Paehtz2012} & $- a + b U^{-1} + c U^{-2}$ \\
      \br      
    \end{tabular}
  \end{indented}
\end{table}

It is interesting to see how the conventional one-species descriptions
emerge from the two-species model upon contraction to a single effective
grain species.  This contraction is clearly not unique but may be
performed in different ways.  On the basis of the two-species flux
balance, discussed in the preceding section, two natural contractions
suggest themselves: one for moderate wind speeds where the flux is
dominated by saltation, and one for strong winds, where the ratio of the
contributions from reptating and saltating grains was found to be
roughly constant.
  
\begin{figure}[t]
  \centering
  \includegraphics[width=0.6\linewidth]{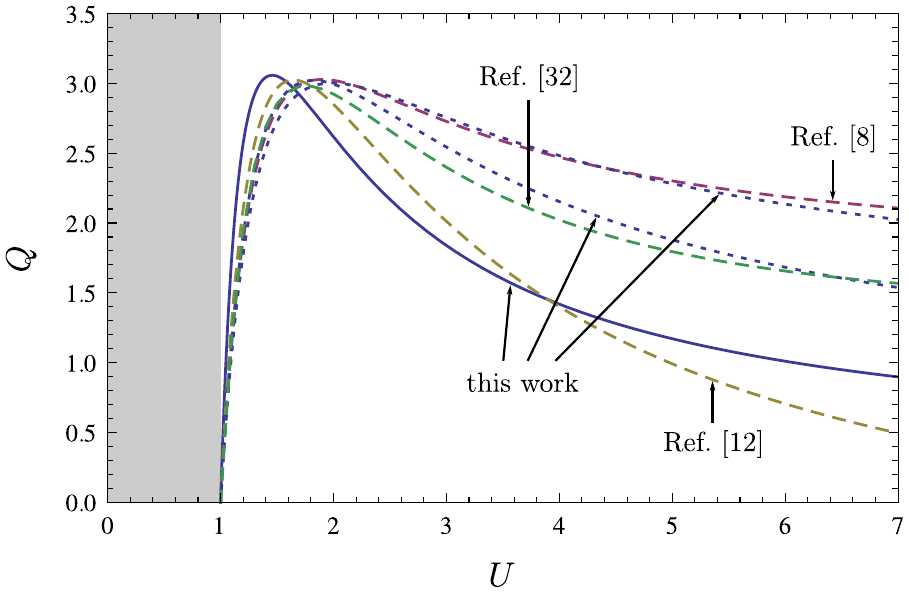}
  \caption{\textbf{The stationary sand-flux laws.}  Predictions from
    the two-species model, equation \eref{eq:transport-law:flux}
    (solid line) with its natural one-species limits
    \eref{eq:transport-law:one-spiecies-limit:onespec-flux} and
    \eref{eq:transport-law:one-spiecies-limit:phi-const} (lower and
    upper dotted line), are compared with transport laws from
    conventional one-species models: Sauermann \etal
    \cite{Sauermann2001} (upper dashed), Dur{\'a}n and Herrmann
    \cite{Duran2006} (lower dashed), and P\"ahtz \etal
    \cite{Paehtz2012} (middle dashed).  All curves have been obtained
    for a grain size of $d=250\,\mathrm{\mu m}$.  The values of the
    free parameters $\alpha$ and $\eta$ in
    \eref{eq:transport-law:flux} were determined by comparison to
    experiments as described in \sref{sec:experiments}.  The
    predictions of the single-trajectory limits of the two-species
    model, \eref{eq:transport-law:one-spiecies-limit:onespec-flux} and
    \eref{eq:transport-law:one-spiecies-limit:phi-const}, were
    adjusted by the parameter rescaling $z^\sal \to 0.18 z^\sal$ and
    $\eta\to 0.138\eta$. }
  \label{fig:fluxlaws}
\end{figure}

The first single trajectory reduction, where the reptating species is
dismissed, is obtained by setting $\varphi=1$ or $\eta=0$, which
corresponds to the flux relation
\begin{equation}
  Q = \left(1-U^{-2}\right)\left( a - b \, U^{-1} \right) \,,
  \label{eq:transport-law:one-spiecies-limit:onespec-flux}
\end{equation}
with a negative coefficient $b<0$ (\ref{sec:app:coeff}).  Since this
equation accounts only for high-energy saltating grains, the resulting
absolute flux is too large compared with the original two-species
description.  This can be corrected by reducing the effective
trajectory height or, in our formalism, by rescaling the effective
height $z^\sal$ at which the fluid drag is balanced with the effective
bed friction.  The lower dotted line in \fref{fig:fluxlaws} was
obtained in this way.

The second single trajectory reduction is motivated by the weak
dependence of the mass fraction $\varphi$ on $\ust$ for strong winds
$\ust \gg \ustt$, which suggests to replace $\varphi$ by its
$\ust$-independent limit $1/(1+\eta/\alpha)$.  This yields
\begin{equation}
  \label{eq:transport-law:one-spiecies-limit:phi-const}
  Q = \left(1-U^{-2}\right) \left( \tilde \alpha_0 + \tilde \beta_0 \, U^{-1} \right) \,,
\end{equation}
with
\begin{equation}
  \label{eq:transport-law:one-species-limit:phi-const-coeff}
  \tilde \alpha_0 = \frac{\alpha_0}{a} \quad  \mathrm{and} \quad
  \tilde \beta_0 = \frac{\beta_0}{2a} + \frac{\tilde \eta}{1+\tilde \eta} \frac{ v^\rep_{z0}}{\ustt} \,,
\end{equation}
and the abbreviation $\tilde \eta \equiv \eta v^\rep/v^\rep_{z0}$.  If
one interprets the (now constant) species mass balance as a free
phenomenological parameter, it may be adjusted by fine-tuning $\eta$,
so that the absolute value of the sand flux predicted by this formula
agrees better with experimental observations.  This is how we obtained
the upper dotted line in \fref{fig:fluxlaws}.

Both \eref{eq:transport-law:one-spiecies-limit:onespec-flux} and
\eref{eq:transport-law:one-spiecies-limit:phi-const} have the functional
form of the transport law proposed by Dur{\'a}n and Herrmann
\cite{Duran2006} based on the focal point assumption.  Note that the
effective height $z^\sal$ is small compared with the saltation height,
where the wind speed obtained from the focal point assumption is very
weak.  As a consequence, the mean drag force is reduced, which results
in a negative parameter $a<0$ in
\eref{eq:transport-law:one-spiecies-limit:onespec-flux} and a negative
sand transport rate $Q<0$ for large shear velocities $U > | a / b |$.
\Tref{tab:transport-law:one-spiecies-limit:fluxlaws} and
\fref{fig:fluxlaws} provide a summary of various stationary aeolian
sand transport laws that have been proposed in the literature, in
comparison with the prediction of our two-species model and its above
single-trajectory contractions.

\subsection{Comparison with experiments}
\label{sec:experiments}
\begin{figure}[t]
  \centering
  \includegraphics[width=0.6\linewidth]{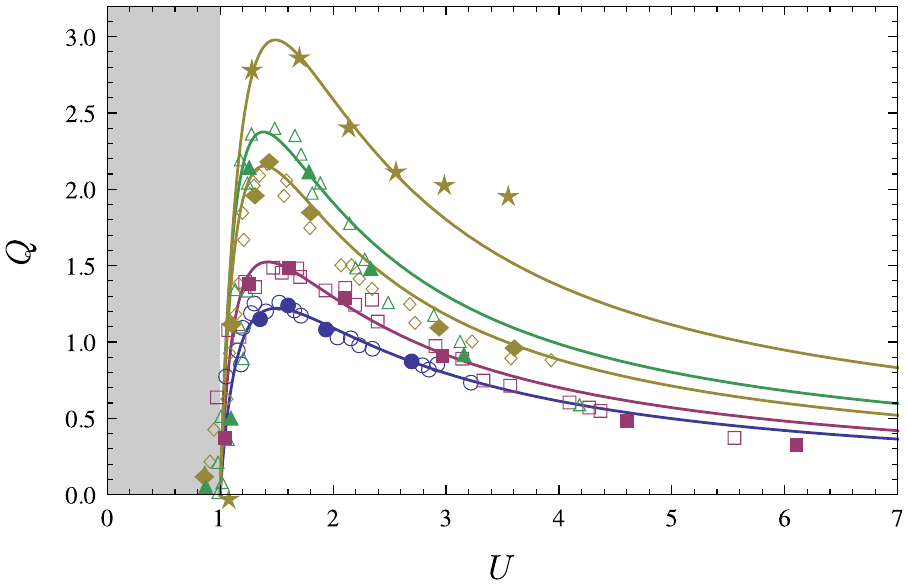}
  \caption{{\bf The rescaled saturated flux $Q=q g / (\varrho_\air
      \ust^3)$ for various grain sizes} ($d=125\,\mathrm{\mu
      m},\,170\,\mathrm{\mu m},\,242\,\mathrm{\mu
      m},\,320\,\mathrm{\mu m},\,242\,\mathrm{\mu m}$ from bottom to
    top): data from wind tunnel measurements using particle tracking
    methods \cite{Creyssels2009} (stars) and sand traps
    \cite{Iversen1999} (other symbols) are compared with the prediction
    by the two-species model, equation \eref{eq:transport-law:flux}
    (solid curves).  The restitution coefficient $\alpha$ served as a
    global but grain-size-dependent fit parameter (see
    \fref{fig:fitparameters}), while $\eta=9$ and $\eta=3.8$ was
    fixed for the sand trap and the particle tracking data,
    respectively. (The open symbols represent several rescaled data
    sets, obtained for a variety of bed slopes.)}
  \label{fig:flux}
\end{figure}

\begin{figure}[t]
  \includegraphics[width=0.48\linewidth]{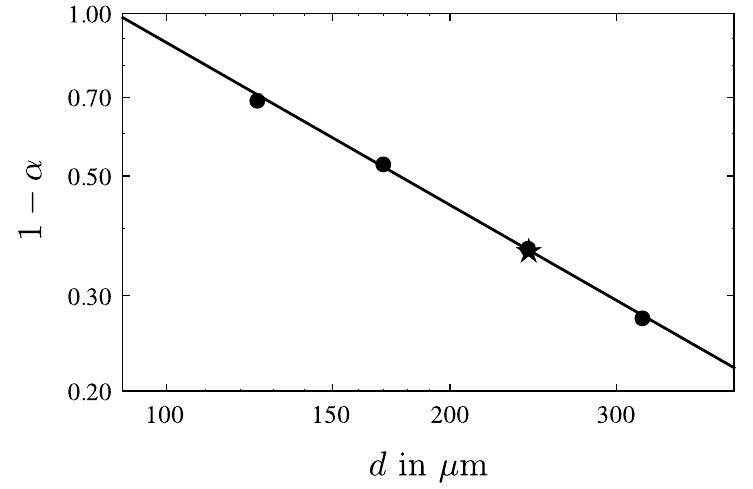}
  \hfill
  \includegraphics[width=0.48\linewidth]{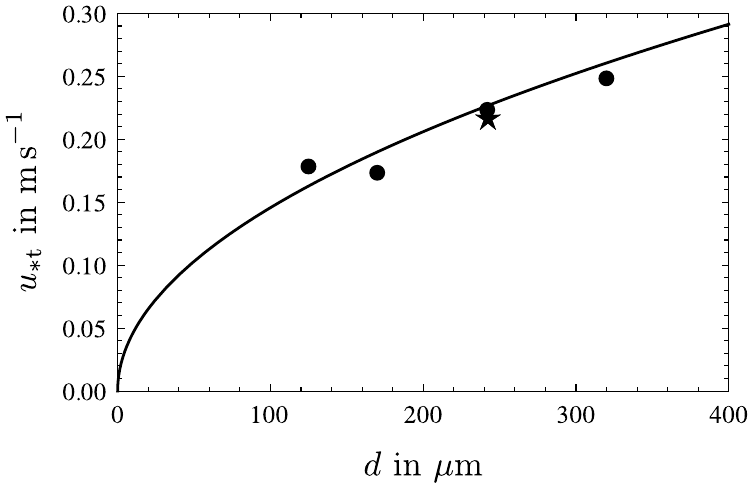}
  \caption{\textbf{The restitution coefficient $\alpha$ (left) and the
      threshold shear velocity $\ustt$ (right) versus the grain size
      $d$.} Parameter values deduced from model fits to flux data in
    \fref{fig:flux}.  The solid lines represent the scaling
    $(d/d_0)^{-1}$ according to \eref{eq:experiments:alpha-d} for $1 -
    \alpha$ and the classical estimate $\ustt = 0.1 \sqrt{\sigma g
      d}$, respectively.}
  \label{fig:fitparameters}
\end{figure}

We now fit our two-species flux relation $Q(U)$ from
\eref{eq:transport-law:flux} to the empirical data from various wind
tunnel measurements \cite{Iversen1999,Creyssels2009}, using $\eta$ and
$\alpha$ as free fit parameters.  As demonstrated in \fref{fig:flux},
we obtain excellent agreement with the experimental data for a
grain-size-independent splash efficiency $\eta$ and an effective
coefficient of restitution $\alpha$ that increases with increasing
grain size $d$. The transport rates for four different grain sizes
collected in \cite{Iversen1999} were obtained using sand traps; the
data in \cite{Creyssels2009} were obtained by particle tracking
techniques for a single grain size only.  The different methods yield
different absolute values for the sand flux, which might be due to
systematic differences in the detection efficiency.  In the model,
these differences are reflected in the (apparently) different
efficiencies with which saltating grains generate reptating grains,
\ie different values of the parameter $\eta$.  We find that $\eta=3.8$
for the data obtained by particle tracking and $\eta=9$ for the sand
trap data, independently of the grain size.  But both data sets were
consistent with the same formula for the bed restitution coefficient
\begin{equation}
  \label{eq:experiments:alpha-d}
  \alpha = 1 - d_0/d \,,
\end{equation}
with $d>d_0=88\,\mathrm{\mu m}$ (left panel of
\fref{fig:fitparameters}).  The rebound becomes increasingly elastic
for larger grains, while $\alpha$ vanishes for smaller grains at a
critical grain diameter $d = d_0$, which is in accord with the
observation that saltation only occurs for sand grains larger than
about $70\,\mathrm{\mu m}$ \cite{shao_book}.  The dependence of the
restitution coefficient $\alpha$ on the grain size is phyically
plausible, for the collision of smaller grains is increasingly
influenced by hydrodynamic interactions (and also by cohesive and
electrostatic forces).  In other words, $d_0$ is a characteristic
grain size marking the transition from a phenomenology typical of sand
to one typical of dust.

Moreover, while we also fine-tune the threshold $\ustt$ for each grain
size by hand when fitting the experimental data, the values used turn
out to be in very good agreement with the expectation $\ustt \approx
0.1\sqrt{\sigma g d}$, as demonstrated by the inset of
\fref{fig:flux}.  Altogether, the stationary sand transport rate
observed in experiments is thus convincingly reproduced by the
two-species result, equation \eref{eq:transport-law:flux}, over a wide
parameter range, with very plausible and physically meaningful values
for the model parameters.

The observed value of $d_0$ can be rationalized by a hydrodynamic
order of magnitude estimate.  Commonly, one relates the crossover from
sand-like to dust-like behaviour to the difference in transport modes
of these two classes of granular.  While sand is transported by grain
hopping dust remains suspended in the air for a while.  Balancing the
settling velocity of the grains by the typical (upward) eddy currents,
which are of the order of $\ustt$, one obtains a minimal sand grain
size of about $1.5 (\nu^2/\sigma g)^{1/3} \approx 30\, \mathrm{\mu m}$
\cite{Bagnold1954}.  To match $d_0=88\, \mathrm{\mu m}$, the eddy
velocity would have to be $4.5 \ustt$.  Two further estimates of $d_0$
may be obtained as follows.  The first assumes that the crossover
marked by $d_0$ is concomitant with the crossover from a more elastic
to a more viscous collision between individual grains.  In this case,
the relevant quantity is the Stokes number, which characterizes the
particle inertia relative to the viscous forces.  Using the
observations of the collision experiments by Gondret \etal
\cite{Gondret1999} for the critical Stokes number $\mathrm{St}$, below
which the impacting particles do not rebound from a rigid wall, one
gets a critical grain size of about $20\, \mathrm{St}^{2/3}
\nu^{2/3}/\sigma/g^{1/3} \approx 20 \, \mathrm{\mu m}$.  To match
$d_0=88\, \mathrm{\mu m}$, the critical Stokes number would have to be
of the order of 190, \ie ten times larger than the value obtained in
\cite{Gondret1999}, for which experimental differences might possibly
be blamed.  Note, however, that the most pertinent argument should be
the one that gives the largest value of $d_0$, as it provides a lower
bound for the grain size contributing to saltation.  A larger estimate
of $d_0$ is indeed obtained by observing that sand grains collide with
the bed, and they are lifted from the bed by turbulent fluctuations in
the wind velocity.  In contrast, suspended and resting dust particles
are protected from bed collisions and turbulent lift, respectively, by
the laminar boundary layer that coats any solid (no slip) boundary.
The characteristic length scale of this laminar coating is related to
the so-called Kolmogorov dissipation scale
$(\nu_\air^3/\varepsilon)^{1/4}$ \cite{Frisch1995}, where
$\varepsilon$ is the dissipation per unit mass of the driving medium
(in our case air at normal conditions).  For the logarithmic wind
profile near the ground, equation \eref{eq:wind:universal}, the
scale-dependent dissipation takes the form $\varepsilon(z) =
\ustt^3/\kappa z$ \cite{Landau1987}.  A self-consistency argument
requiring that the Kolmogorov dissipation scale at height $z$ above
the ground equals $z$ (if no externally imposed roughness scale is
available) yields $z \simeq [\nu_\air^3/\varepsilon(d_0)]^{1/4}$.
Extrapolating the logarithmic profile with $\ustt = 0.1 \sqrt{\sigma g
  d}$ down to the scale $z$, we obtain $z = 4.6 (\nu_\air^2/\sigma
g)^{1/3} \approx 10^2\, \mathrm{\mu m}$.  We find this estimate to be
in excellent agreement with the data in the left panel of
\fref{fig:fitparameters}, which suggests
\begin{equation}
  \label{eq:experiments:d0}
  d_0 \approx 4  (\nu_\air^2/\sigma g)^{1/3}  \,.
\end{equation}

\section{Conclusions}
\label{sec:conclusions}
We have developed a ``second generation'' continuum description of
aeolian sand transport, relaxing the strong mean-field approximation
inherent in the classical single-trajectory models such as
\cite{Sauermann2001}, and introducing a two-species framework, as
advocated in \cite{Andreotti2004}.  We started from the physics on the
grain scale and corroborated our explicit analytical expressions by a
comprehensive comparison with empirical data for the splash, the wind
velocity field and the sand flux, covering a broad range of ambient
conditions and grain sizes.  While the usefulness of our general
approach of contracting the complicated splash and saltation process
into a drastically reduced two-species picture was strongly suggested
by empirical and theoretical work, it necessarily involved some free
phenomenological coefficients.  For our two-species model, these are,
apart from a few minor numerical coefficients listed in
\tref{tab:parameters}, the effective bed restitution coefficient
$\alpha(d)$ for the saltating grains of diameter $d$ and the parameter
$\eta$ that fixes the number of reptating grains ejected by the
average impacting saltating grain.  Both parameters were found to take
consistent and plausible values if used as free parameters when
fitting the empirical flux data obtained in wind tunnel experiments.
In particular, the size dependence of the restitution coefficient fits
perfectly to the phenomenological criteria commonly used to
distinguish dust from sand.  The predicted two-species stationary flux
law \eref{eq:transport-law:flux} was found to be in excellent
agreement with comprehensive data from different sources.  It should
provide an excellent analytical starting point for a variety of
advanced applications calling for a more faithful description of the
saltation process so far available---from wind-driven structure
formation in the desert to saltation-driven dust production and
emission.

\ack
\label{sec:acknowledgments}
We thank Thomas P\"ahtz and Jasper Kok for inspiring discussions and
for making the manuscript of \cite{Paehtz2012} available to us prior
to its publication.

\appendix

\section{Solving the Prandtl turbulent closure}
\label{sec:app:closure}

With \eref{eq:basics:stressdecomp} and the assumption of an
exponential decay of the grain-borne shear stress with height,
equation \eref{eq:wind:taug-vs-z}, the modified Prandtl turbulence
closure \eref{eq:wind:turbulence-closure} reads as
\begin{equation}
  \label{eq:app:closure}
  \partial_z u = \frac{\ust}{\kappa z} \sqrt{1-\frac{\taug(0)}{\tau}
  \rme^{-z/z_\rmm}} \,.
\end{equation}
To make analytical progress,  we follow S{\o}rensen
\cite{Sorensen2004} in approximating the square root by a secant.  The
right-hand side of \eref{eq:app:closure} has the functional
form $\sqrt{1 - \epsilon x}$,
with $x \equiv \exp(-z/z_\rmm)$ and $\epsilon \equiv \taug(0)/\tau$, 
which we approximate by a secant of the form $a x + b
(1-x)$. Matching the points $\{0,1\}$ and $\{1,\sqrt{1-\epsilon}\}$
yields $b=1$ and $a=\sqrt{1-\epsilon}$, hence
\begin{equation}
  \label{eq:squarroot-approx}
  \sqrt{1 - \epsilon x} \approx 1- (1- \sqrt{1-\epsilon})x \,.
\end{equation}
Under stationary conditions, using \eref{eq:basics:taug-ground},
we have $\sqrt{1-\epsilon}=\sqrt{(\tau-\taug(0))/\tau}=\ustt/\ust$,
which leads to \eref{eq:wind:closure-approx}.  This avoids an
artefact of S{\o}rensen's original approximation \cite{Sorensen2004},
$\sqrt{1 - \epsilon x} \approx 1 - \epsilon x$, which implies
$\sqrt{\taua(0)/\tau} = 1 - \taug(0)/\tau$ for the shear stress at the
ground, as already criticized by Dur{\'a}n and Herrmann
\cite{Duran2006}.

\section{The two-species approach for the wind profile}
\label{sec:app:2spec-wind}

In \eref{eq:basics:grainstress-div} of \sref{sec:basics}, the
grain-borne shear stress was split up into the contributions from
saltating and reptating particles, respectively.  The ratio of the
grain-borne shear stresses at the ground,
\begin{equation}
  \label{eq:app:2spec-wind:taug0-ratio}
  \frac{\taug^\rep(0)}{\taug^\sal(0)} = \alpha \frac{ v^\rep}{v^\rep_{z0}} \frac{1-\varphi}{\varphi} \equiv \tilde \alpha \, \frac{1-\varphi}{\varphi} \,,
\end{equation}
immediately follows from
\eref{eq:basics:taug-ground-parabola} and
\eref{eq:basics:alpha}, and from our simplifying assumption that
the reptating grains are ejected vertically. Assuming the exponential
decay of the grain-borne shear stress $\taug$ with height to hold for
both components, the Prandtl turbulent closure reads
\begin{equation}
  \label{eq:app:2spec-wind:closure}
  \rho_\air \kappa^2 z^2 (\partial_z u)^2  = \tau - \taug^\rep(0) \rme^{-z/
    z^\rep_\rmm}  - \taug^\sal(0) \rme^{-z/ z^\sal_\rmm} \,.
\end{equation}
We exploit the strong scale separation $z^\sal_\rmm/z^\rep_\rmm \simeq
10^{2}$ between the characteristic jump heights of saltating and
reptating grains \cite{Bagnold1954,Andreotti2010}, on which the
two-species model is based. (The precise value turns out to be
irrelevant to our discussion.)  It allows to solve the closure for
two separate height ranges:
\begin{inparaenum}[(i)]
\item $z <z^\rep_\rmm \ll z^\sal_\rmm$ associated with reptation,
 where we may set $\exp(-z/ z^\sal_\rmm) \to 1$, and
\item $z \gg z^\rep_\rmm$, associated with saltation, where we may set
  $\exp(-z/ z^\rep_\rmm) \to 0$.
\end{inparaenum}
Applying the secant approximation for the square root as described in
\ref{sec:app:closure}, we can perform the integrations within
both ranges and match the asymptotic solutions at $z = z^\rep_\rmm$.
Using \eref{eq:basics:taug-ground} and
\eref{eq:app:2spec-wind:taug0-ratio} to eliminate $\taug^\sal(0)$ and
$\taug^\rep(0)$, this yields
\begin{equation}
  \label{eq:app:2spec-wind:windprofile}
  \fl
  u(z) = \cases{
    \frac{S \ust}{\kappa} \ln\frac{z}{z_0} - \frac{S \ust - \ustt}{\kappa}\left[ \Ei\!\left(z_0/z^\rep_\rmm\right) - \Ei\!\left(z/z^\rep_\rmm\right) \right] \,, & $z \le z^\rep_\rmm$ \\
    u(z^\rep_\rmm) + \frac{\ust}{\kappa}
    \ln\frac{z}{z^\rep_\rmm} \\ \qquad \quad- \frac{\ust (1 - S )}{\kappa} \left[ \Ei\!\left(z^\rep_\rmm/z^\sal_\rmm\right) -
      \Ei\!\left(z/z^\sal_\rmm\right) \right] \,, & $z > z^\rep_\rmm \,,$
  }
\end{equation}
where we abbreviated
\begin{equation}
  \label{eq:app:2spec-wind:windprofile-abbreviations}
  S = \sqrt{1 - \frac{(1-\taut/\tau)\varphi}{\varphi + (1-\varphi)\tilde
      \alpha}} \,.
\end{equation}
\begin{figure}[t]
  \centering
  \includegraphics[width=0.6\linewidth]{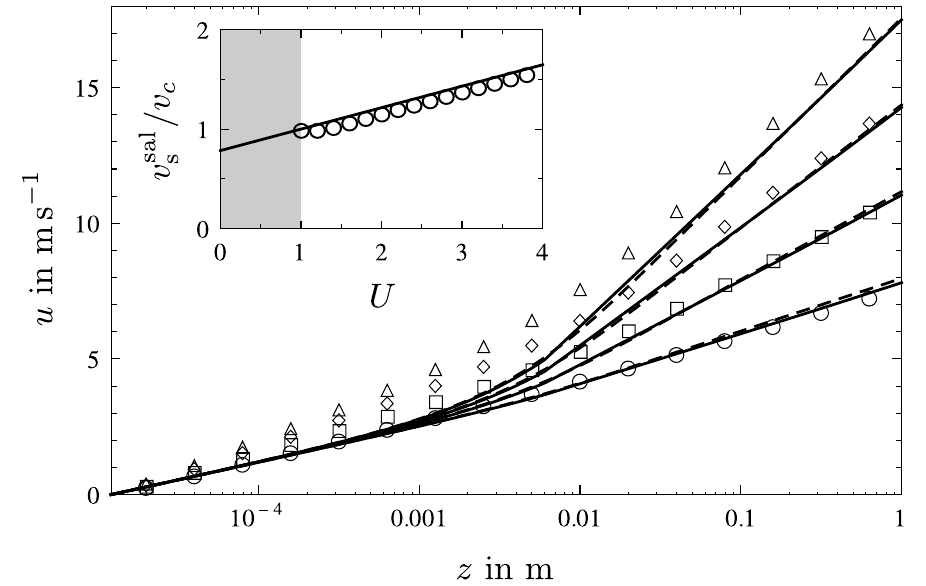}
  \caption{Comparison of the wind velocity profile
    \eref{eq:app:2spec-wind:windprofile} obtained from the
    self-consistent numerical solution of
    \eref{eq:app:2spec-wind:selfconsitent-velocity} (solid lines) with
    the pre-averaged profile \eref{eq:wind:windprofile} (dashed lines)
    for grains of diameter $d=242\,\mathrm{\mu m}$ and for various shear
    velocities ($U=1.5$, 2.5, 3.5 and 4.5 from bottom to top).  As
    explained in the main text, both approaches coincide for $z_\rmm =
    z^\rep_\rmm$.  Additionally, we show the results from numerical
    solutions of the turbulence closure \eref{eq:app:2spec-wind:closure}
    (open symbols) to support the approximate expression
    \eref{eq:app:2spec-wind:windprofile}.  \emph{Inset:} The saltation
    velocity $v^\sal$, rescaled by the minimum saltation velocity
    $v^\sal_\rmc$ needed to eject any grains, over the shear velocity:
    self-consistent numerical solution of
    \eref{eq:app:2spec-wind:selfconsitent-velocity} (symbols) and its
    analytical prediction in the limit $(1-\taut/\tau)\varphi \to 0$,
    for which $u(z)$ is given by
    \eref{eq:app:2spec-wind:windprofile-phizero}, (line).}
  \label{fig:app:2spec-wind:windprofile}
\end{figure}
Using this in the force balance, \eref{eq:kinetics:saltation-velocity},
we gain the implicit equation
\begin{equation}
  \label{eq:app:2spec-wind:selfconsitent-velocity}
  v^\sal = u(z^\sal,v^\sal) - v^\sal_\infty \,,
\end{equation}
where the fraction $\varphi$ of saltating grains itself depends on the
velocity $v^\sal$ of the saltating grains via
\eref{eq:basics:weightfct}.  To solve it, we note that the fraction
$\varphi$ of saltating particles is a decreasing function of the shear
stress $\tau$, bracketed by 1 and 0 (\fref{fig:species-fluxes}, right
panel).  Hence, the product $(1-\taut/\tau) \varphi$ is a small number
for all $\tau \ge \taut$. Setting it to zero in
\eref{eq:app:2spec-wind:windprofile}, we obtain
\begin{equation}
  \label{eq:app:2spec-wind:windprofile-phizero}
  \fl
  u(z) \sim \frac{\ust}{\kappa} \ln\left(\frac{z}{z_0}\right) - \frac{\ust-\ustt}{\kappa} \times
  \cases{[\Ei(z_0/z^\rep_\rmm)-\Ei(z/ z^\rep_\rmm)] \, , & $z \le z^\rep_\rmm$ \\ [\Ei(z_0/z^\rep_\rmm) - \Ei(1)] \, , & $z > z^\rep_\rmm$}
\end{equation}
Since the exponential integral $\Ei(z / z_\rmm)$ vanishes rapidly for
increasing $z> z_\rmm$, we recover the wind speed profile
\eref{eq:wind:windprofile} successfully employed in the one-species
models, with the characteristic decay height given by $z_\rmm \equiv
z^\rep_\rmm$.  Inserting \eref{eq:app:2spec-wind:windprofile-phizero}
into \eref{eq:app:2spec-wind:selfconsitent-velocity} yields an affine
increase of $v^\sal$ with $\ust$, in good accord with the exact
numerical solution of \eref{eq:app:2spec-wind:selfconsitent-velocity}
(\fref{fig:app:2spec-wind:windprofile}, right panel).

For the numerical solution, one has to deal with the two free
parameters $z^\rep_\rmm$ and $\alpha$.  While the former is directly
related to the wind profile, the latter is a fit parameter of the
model, determined from a comparison of the predicted sand transport
rate with empirical data (\fref{fig:flux}).  To make progress, we vary
$z^\rep_\rmm$ and take the value of $\alpha$ from
\sref{sec:transport-law}, where the sand flux is estimated by means of
the pre-averaged approach for the wind profile.  For a grain diameter
$d=242\,\mathrm{\mu m}$, we obtained $\alpha \approx 0.63$ (\ie
$\tilde\alpha=1.4$), which is consistent with collision experiments,
as argued in \cite{Duran2006}.  From the numerical solution of
\eref{eq:app:2spec-wind:selfconsitent-velocity}, we find good
agreement between the self-consistently gained $u(z)$ and wind tunnel
measurements \cite{Rasmussen2008} for $z^\rep_\rmm=25 d$, which is
exactly the same value as obtained in \sref{sec:wind} within the
pre-averaged approach.  This supports our observation that
\eref{eq:app:2spec-wind:windprofile} can be approximated by taking the
limit $\varphi \to 0$ in \eref{eq:app:2spec-wind:windprofile-phizero},
which yields the pre-averaged wind profile for $z_\rmm\equiv
z^\rep_\rmm$.  This result is almost independent of the ratio
$z^\sal_\rmm/z^\rep_\rmm$, for
\eref{eq:app:2spec-wind:windprofile-phizero} is independent of
$z^\sal_\rmm$.  Thereby, we formally confirm the intuitive expectation
that the feedback of the grains on the wind profile is predominantly
due to the many reptating particles and hardly affected by the
saltating particles.

\section{Reptation velocity}
\label{sec:app:reptation-velocity}
\begin{figure}[t]
  \centering
  \includegraphics[width=0.6\linewidth]{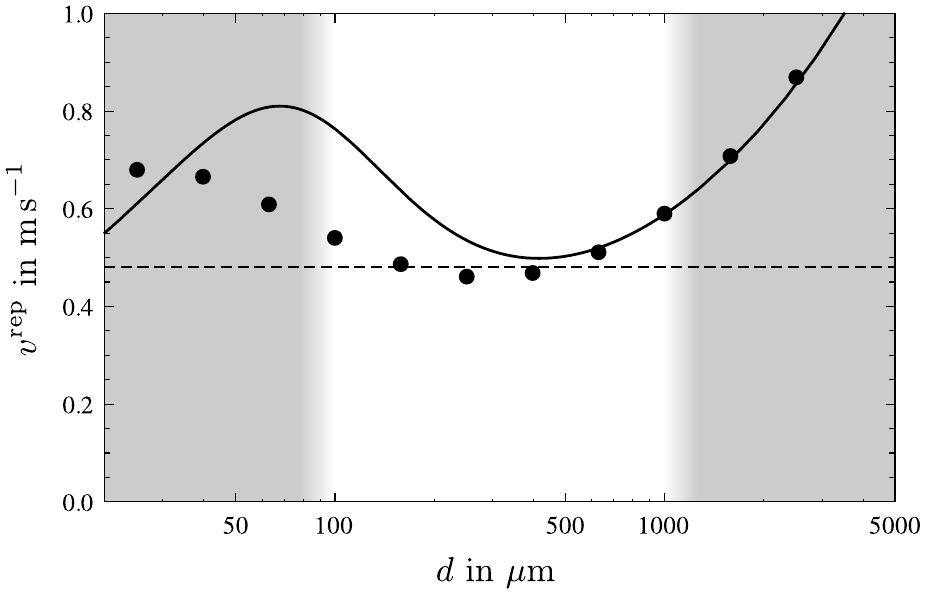}
  \caption{The Reptation velocity $v^\rep$ against the grain diameter
    $d$. The dots correspond to the numerical solution of
    \eref{eq:app:reptation-velocity:reptation-force} averaged
    over one trajectory, while the solid and dashed lines represent
    the approximate solution
    \eref{eq:app:reptation-velocity:approx-velocity} and the
    representative value $v^\rep = 0.7 (\sigma \nu_\air g)^{1/3}
    \approx 0.5\;\mathrm{m\,s^{-1}}$ proposed in
    \eref{eq:kinetics:reptation-velocity}, respectively.  The
    latter can be understood as an average over the relevant grain
    sizes (white background).}
  \label{fig:app:reptation-velocity}
\end{figure}

As to the saltating grain fraction in \ref{sec:kinetics}, we estimate
the transport velocity of the reptating grains from the grain-scale
physics, \ie from the hop-averaged horizontal velocity of an
individual grain.  (Note that we do not introduce a new variable to
distinguish the grain-scale velocity from the mean transport velocity,
because the context prevents confusion.)  The time-dependent velocity
of a reptating grain obeys the drag relation
\begin{equation}
  \label{eq:app:reptation-velocity:reptation-force}
  \partial_t v^\rep = \frac{g}{(v^\rep_\infty)^2}\, \left| u-v^\rep \right| \, \left(u-v^\rep\right) \,,
\end{equation}
similar to that for saltating grains \cite{Sauermann2001,Duran2006}.
For saltating grains, an additional friction force (besides the drag
force) would appear on the right-hand site of the equation of motion,
representing the mean loss of momentum upon rebound.  But, since the
reptating grains perform only a single hop, such a friction term does
not enter \eref{eq:app:reptation-velocity:reptation-force}.  We assume
that the ejection is essentially vertical with the initial velocity
$v^\rep_{z0}$ of the order of $\ustt$.  A more accurate discussion
would not substantially change our findings, as confirmed by the
numerical solution of
\eref{eq:app:reptation-velocity:reptation-force}.  Note that the
nonlinearity of the drag law entails a time-dependent ``terminal
settling velocity'' $v^\rep_\infty$, dependent on the actual relative
grain velocity $u-v^\rep$ (e.g., \cite{Jimenez2003}).  However, for our
purpose, and in view of the low reptation trajectories, we can safely
approximate the reptation velocity from
\eref{eq:app:reptation-velocity:reptation-force} by inserting the wind
speed $u(z^\rep)$ at a given reptation height and the steady-state
terminal velocity
\begin{equation}
  \label{eq:app:reptation-velocity:terminal-velocity}
  v^\rep_\infty = \sqrt{\sigma g d} \left[0.95 + 20 \, \nu_\air /\sqrt{\sigma g
      d^3} \right]^{-1}
\end{equation}
derived from the effective drag law proposed in \cite{Jimenez2003},
similar to \eref{eq:kinetics:terminal-velocity} (see also
\cite{Sauermann2001,Duran2006}).  Neglecting moreover vertical drag
forces, the maximum height of the reptation trajectory is
$(v^\rep_{z0})^2/(2 g) \approx 10 d < z_\rmm $.  Consequently, we may
insert the ground-level wind field \eref{eq:wind:universal} and obtain
the mean reptation velocity
\begin{equation}
  \label{eq:app:reptation-velocity:approx-velocity}
  v^\rep \approx \frac{\ustt \ln(z^\rep/z_0)}{\kappa \left[ 1 + \kappa \; {(v^\rep_\infty)}^2/\left(2v^\rep_{z0} \ustt  \ln(z^\rep/z_0)\right) \right]} \,,
\end{equation}
where we approximated the time of flight by that for a parabolic
trajectory, $2 v^\rep_{z0} /g$, as usual.  Inserting the empirical
observation $v^\rep_{z0} \approx \ustt$ as well as $\ustt^2=0.01\sigma
g d$ and $z_0 = d/20$, see \tref{tab:parameters}, results in an
estimate for the reptation velocity $v^\rep$ as a function of the
grain diameter $d$, illustrated by a solid line in
\fref{fig:app:reptation-velocity}, which we may identify with the
(mean field) transport velocity of the reptating grain fraction.

The yet undetermined value of the effective reptation height $z^\rep$
is expected to be of the order of the maximum height
$(v^\rep_{z0})^2/(2g)$ of the trajectory.  Indeed, if we compare the
approximate result given by
\eref{eq:app:reptation-velocity:approx-velocity} with the
numerical solution $v^\rep(t)$ of
\eref{eq:app:reptation-velocity:reptation-force} averaged over
the whole trajectory,
\begin{equation}
  \label{eq:app:reptation-velocity:avrg-velocity}
  \overline{v^\rep} = \frac{g}{2 v^\rep_{z0}} \int_0^{2v^\rep_{z0}/g} \rmd t\; v^\rep(t) \,,
\end{equation}
we obtain a good match for large grain diameters $d>800\; \mathrm{\mu
  m}$ for $z^\rep =0.14 \ustt^2/(2g)$.  Note, however, that the
numerical solution yields an almost $d$-independent reptation velocity
$\overline{v^\rep} \approx 0.5 \; \mathrm{m \,s^{-1}}$ for the
relevant grain sizes $d \approx 100 \, \mathrm{\mu m} \dots 1 \,
\mathrm{mm} $, as illustrated in
\fref{fig:app:reptation-velocity}.  To find an estimate for
$\overline{v^\rep}$, which still captures the dependence on other
parameters appearing in
\eref{eq:app:reptation-velocity:approx-velocity}, we evaluate
this equation for a representative grain diameter within the relevant
range.  A closer look at
\eref{eq:app:reptation-velocity:approx-velocity} reveals that the
reptation velocity at the inflection point $d \approx 6.7
\nu_\air^{2/3} (\sigma g)^{-1/3} \approx 150 \; \mathrm{\mu m}$ can be
approximated by $v^\rep \approx (\nu_\air \sigma g)^{1/3} = 0.68 \;
\mathrm{m \,s^{-1}}$, which provides the wanted parameter dependence.
To better match the absolute values found from the numerical solution,
we insert a factor $0.7$ by hand, thus arriving at the analytical
estimate given in \eref{eq:kinetics:reptation-velocity}.

\section{The coefficients of the transport law}
\label{sec:app:coeff}
\begin{table}[b]
  \caption{The Parameters occurring in equations \eref{eq:app:coeff:a}
    to \eref{eq:app:coeff:gamma0} which are either numerical constants
    or dependent on the grain diameter $d$, the gravitational
    acceleration $g=9.81\;\mathrm{m\, s^{-1}}$, the sand--air density
    ratio $\sigma=2163$, or the kinematic viscosity of air, $\nu_\air =1.5 \times 10^{-5}\;\mathrm{m^2\,s^{-1}}$.}
  \label{tab:parameters}
  \begin{indented}
  \item[] \begin{tabular}{@{}ll} \br
      parameter & value or formula \\
      \mr
      $\ustt$ & $0.1 \sqrt{\sigma g d}$\\
      $\kappa$ & $0.4$ \\
      $\eta$ & $9$ \cite{Iversen1999}, $3.8$ \cite{Creyssels2009}\\
      $v^\sal_\rmc$ & $10 \ustt$  \\
      $v^\rep_{z0}$ & $\ustt$ \\
      $z_0$ & $d/20$ \\
      $z_\rmm$ & $25\,d$ \\
      $\alpha$ & $1-d_0/d$, \quad $d_0 = 4  (\nu_\air^2/\sigma g)^{1/3}$ \\
      $v^\sal_\infty$ & $\sqrt{\sigma g d/\alpha} \left[1.3 + 41 \sqrt{\alpha} \, \nu_\air  /\sqrt{\sigma g d^3} \right]^{-1}$ \\
      $v^\rep$ & $0.7 (\sigma \nu_\air g)^{1/3}$ \\
      \br
    \end{tabular}
  \end{indented}
\end{table}

Here we give explicit expressions for the coefficients occurring in
the saturated sand flux \eref{eq:transport-law:flux}:

\begin{eqnarray}
  \label{eq:app:coeff:a}
  a &= 2\alpha (1 + \tilde \eta) \frac{v^\sal_\rmc +
    v^\sal_\infty}{\ustt} - 2\alpha \frac{1+\tilde
    \eta}{\kappa}\mathcal{E} \! \left[ z_0\exp \!
    \left( \kappa \frac{v^\sal_\rmc + v^\sal_\infty}{\ustt}\right)
  \right] \\
  \label{eq:app:coeff:b}
    b &= a - 2 \alpha \frac{v^\sal_\rmc}{\ustt}
\end{eqnarray}
and
\begin{eqnarray}
    \label{eq:app:coeff:alpha0}
    \alpha_0 &= \frac{a^2}{(1 + \tilde \eta)^2} \\
    \label{eq:app:coeff:beta0}
    \beta_0 &= \frac{2 \tilde \eta a}{1 + \tilde \eta} \left( \frac{v^\rep_{z0}}{\ustt} + \frac{a}{1 + \tilde \eta}  - \frac{b}{\tilde \eta} \right) \\
    \label{eq:app:coeff:gamma0}
    \gamma_0 &= \frac{\tilde \eta v^\rep_{z0}}{\ustt}
    \left[\frac{\tilde \eta v^\rep_{z0}}{\ustt} + 2(a-b) \right] -
    \left( \frac{1 + \tilde
        \eta}{2a} \beta_0 \right)^2 
  \end{eqnarray}
with the abbreviations
\begin{equation}
  \label{eq:app:coeff:fracte}
  \mathcal{E}(z) = \Ei(z_0/z_\rmm) - \Ei(z/z_\rmm)
\end{equation}
and
\begin{equation}
  \label{eq:app:coeff:alphatilde}
  \tilde \eta \equiv \eta \, v^\rep/v^\rep_{z0} \,.
\end{equation}
All parameters occurring in these equations are listed in
\tref{tab:parameters}.  As explained in the main text, these
quantities are determined as follows.  From collision experiments, we
obtain the numerical values of the critical impact velocity
$v^\sal_\rmc$ and the vertical ejection speed $v^\rep_{z0}$.  The
roughness length $z_0$ and the height $z_\rmm$ (the characteristic
decay height of the air-borne sand density) are estimated by fitting
experimentally observed wind velocity profiles above the saltation
layer.  Finally, the parameters $\eta$, $\alpha$ and the threshold
shear velocity $\ustt$ are determined by fitting the sand transport
rate to wind tunnel measurements.

\newpage
\section*{References}
\bibliographystyle{iopart-num} 
\bibliography{2spec}
\end{document}